\documentclass[pra, twocolumn]{revtex4-1}
\usepackage{amsmath}
\usepackage{amsfonts}
\usepackage{amssymb}
\usepackage{graphicx}
\usepackage{dcolumn}
\usepackage{bm}

\begin{document}

\preprint{APS/123-QED}

\title{Nonlinear Coherent Perfect Absorption}
\author{K. Nireekshan Reddy}
\author{S. Dutta Gupta}
\altaffiliation{sdghyderabad@gmail.com}
\affiliation{School of Physics, University of Hyderabad, Hyderabad-500046, India}

\date{\today}

\begin{abstract}
We review some of the recent concepts and their realization exploiting the perfect destructive interference of light in micro and nano structures. One refers to optical structures where the effective absorption can be controlled and maximized to perfect absorption. The reported effects depend crucially on the coherent nature of the exciting radiation. Achieved with a single (two or more) incident plane wave (waves) the effect carries the name of critical coupling (coherent perfect absorption). Thus in a system supporting critical coupling (CC) or coherent perfect absorption (CPA) all the incident radiation can be absorbed leading to null scattering. In particular all the incident light energy can be channeled into a specified mode of a multimodal structure if such modes are supported by the system. We present a brief overview of CC and CPA in linear systems to recount their underlying concepts as time-reversed lasing and some of their futuristic applications. Next we review our work on the nonlinear extensions of CC and CPA where one or more of the layered media could be nonlinear with Kerr-type nonlinearity. The dispersive nonlinearity is shown to offer a practical handle over the process of perfect absorption by incident laser power. Further we show that the nonlinear periodic structure can support gap solitons which absorbs all the incident energy and do not scatter any light outside the hetero-guide.
\end{abstract}

\maketitle
In recent years there has been a great deal of interest to exploit one of the fundamental properties of light, namely, its ability to interfere, to engineer the absorption in a given structure. In a series of papers \cite{Wan-Sc,trl}, destructive and constructive interference with two coherent inputs were shown to lead to a perfectly absorbing or a scattering structure. The original idea stems back to several decades back when Frederick showed that two incoming waves falling on a metallic grating can render it to be perfectly absorbing \cite{frederick}. There was another interesting proposal by Yariv where back scattered reflected waves were ruled out in a coupled fiber-disk resonator system \cite{yariv-cc1}. Experimental realization came from a different group which clearly showed how to engineer the absorptive properties in a system \cite{Vah}. There were several reports on planar geometries using absorptive polymers or metal dielectric composites on such systems, wherein, all the incident radiation is absorbed by the structure \cite{Bulovic, SDG-CC}. This phenomenon of total absorption with a single incident beam is now referred to as critical coupling (CC). With two coherent inputs the same physical phenomenon of total absorption is called coherent perfect absorption (CPA) \cite{Wan-Sc,opex-12}. Incidentally the standard laser near threshold is a time-reversed analog of CPA. There have been recent efforts to combine the notions of CPA and parity-time ($\mathcal{PT}$) symmetry \cite{longhi-cpa-pt1}. As of now linear properties of CC and CPA are well understood. In view of the lack of the studies on nonlinear systems, a detailed study was launched by Reddy \textit{et. al.}, to investigate the role of nonlinearity as an additional handle \cite{knr1,knr2,knr3}. Note that, CC and CPA hinges on a very delicate balance of phase and amplitude relations. A nonlinear counterpart offers the remarkable possibility to achieve CC or CPA resonances by means of adjusting the laser power. In this review we present a brief overview of what has been achieved so far on the linear and nonlinear systems exhibiting perfect absorption and super-scattering. The goal is to expose to the reader the wide variety of possibilities and the potential applications in diverse areas of optics.
\par%
The organization of the review is as follows. In Section~\ref{S1}, we review critical coupling and coherent perfect absorption in linear systems. We then investigate CC in Kerr nonlinear structures in Section~\ref{S2}. In Section~\ref{S3}, we extend out studies to understand CPA in Kerr nonlinear slab. Gap solitons in the context of CPA in a nonlinear periodic structure are studied in Section.~\ref{S4}. In conclusions we  summarize the important results in Section~\ref{S5}
\section{Linear coherent perfect absorption}\label{S1}
\subsection{Critical Coupling}\label{S1-1}
Critical coupling (CC) refers to a case when all the incident radiation on micro- or nano- surfaces is completely absorbed without any scattering. The coupling of optical power between micro-resonators and dielectric waveguides was considered by Yariv theoretically \cite{yariv-cc1}. It was shown that when the internal losses of the resonator are equal to the coupling losses (with waveguide), one has null transmission from the structure. This was attributed to the destructive interference of the interfering waves. It was also shown that this null transmission doesn't depend on type of coupling and the resonator. This was experimentally realized in a fiber taper to a silica-micro-sphere system by Cai \textit{el. al.} \cite{Vah}. 
\begin{figure}[h]
\center{\includegraphics[width=6cm]{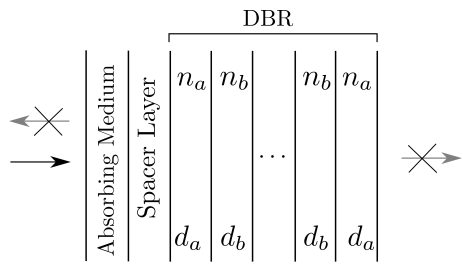}}
\caption{Schematics of the critical coupling structure in planar geometery and illumination.} 
\label{fig:1}
\end{figure}
Later, this was extended to a planar geometry with the help of layered stratified media \cite{Bulovic}. As shown in Fig.~\ref{fig:1}, the planar geometry consisted of a very thin (5 nm) absorbing medium (organic dye), spacer layer and a distributed Bragg reflector (DBR). The DBR was chosen to ensure null transmission from the structure when the wavelength of incident (at normal incidence) radiation falls in the stopgap. The thickness of the spacer layer was tuned so as to ensure the perfect destructive inference among all the reflected waves demanding equality in amplitudes and $(2n+1)\pi$ offset in phase. Thus, one has no reflection and transmission from the structure, leading to perfect absorption of light. In such a situation one says that light is critically coupled to the structure. The wavelength at which CC occurs depends mainly on the properties of the absorber layer. For example, given an absorber like organic dye \cite{Bulovic}, one cannot really have a good control over the location of CC dip. Note that heterogeneous media can offer a true handle over the absorptive properties of the materials (both in linear and nonlinear regimes) \cite{Bohren,tmat}. It was this property of  the metal-dielectric composite medium, which was exploited to demonstrate the flexibility over the location of CC \cite{SDG-CC}. The volume fraction of metal inclusions offers the needed flexibility to control the location of localized plasmon resonance and the oscillator strength. It was also shown that for higher volume fraction of metal inclusions (higher oscillator strengths) one can even have CC at two different wavelengths. Similar ideas were also extended to oblique incidence case  and CC was achieved for both TE and TM polarizations \cite{cc-oblique}. It was also shown that for high oscillator strength one can excite longitudinal bulk plasmons with TM polarized light which are absent in TE case. The reflecting properties of the meta-materials were also used to study CC in Fabry-Perot geometry \cite{sdg-cc-mm}. CC was also demonstrated experimentally in plasmonic cavity arrays \cite{cc-sp-ar}.
\par%
Apart from the fundamental studies, there have been proposals and demonstrations to realize applications of CC. It was shown that by controlling the coupling between the ring resonator and optical waveguide, one can realize switches, which operate at very low thresholds \cite{yariv-cc2}. An experimental demonstration showed that one can have a good control over the transmitted power in the fiber by adjusting the internal losses of the ring resonator near CC resonance. The wavelength selective optical amplification and oscillations were also demonstrated by other group \cite{yariv-cc3}. Since all incident energy is stored in the structure at CC, there will be large buildup of fields in the absorbing layer and one can mimic cavity QED in strong coupling regime \cite{cc-qed,sdg-cc-mm}.
\subsection{Coherent perfect absorption}\label{S1-2}
As noted in the Sec.~\ref{S1-1}, CC in the planar structures involves unidirectional illumination geometry and perfect destructive interference among all the reflected waves. A relatively much simpler geometry was proposed by Chong \textit{et. al,} to study such destructive-interference-assisted absorption known as coherent perfect absorption (CPA) \cite{Wan-Sc}. CPA geometry consisted of a silicon wafer, illuminated at normal incidence from both the sides. It was shown that absorption of the medium can be modulated by changing the phase difference between the two incident beams. Later, this was extended to the oblique incidence case \cite{opex-12}. As noted earlier the absorbing slab was again chosen to be a metal-dielectric composite to have greater flexibility. 
\begin{figure}[h]
\center{\includegraphics[width=8cm]{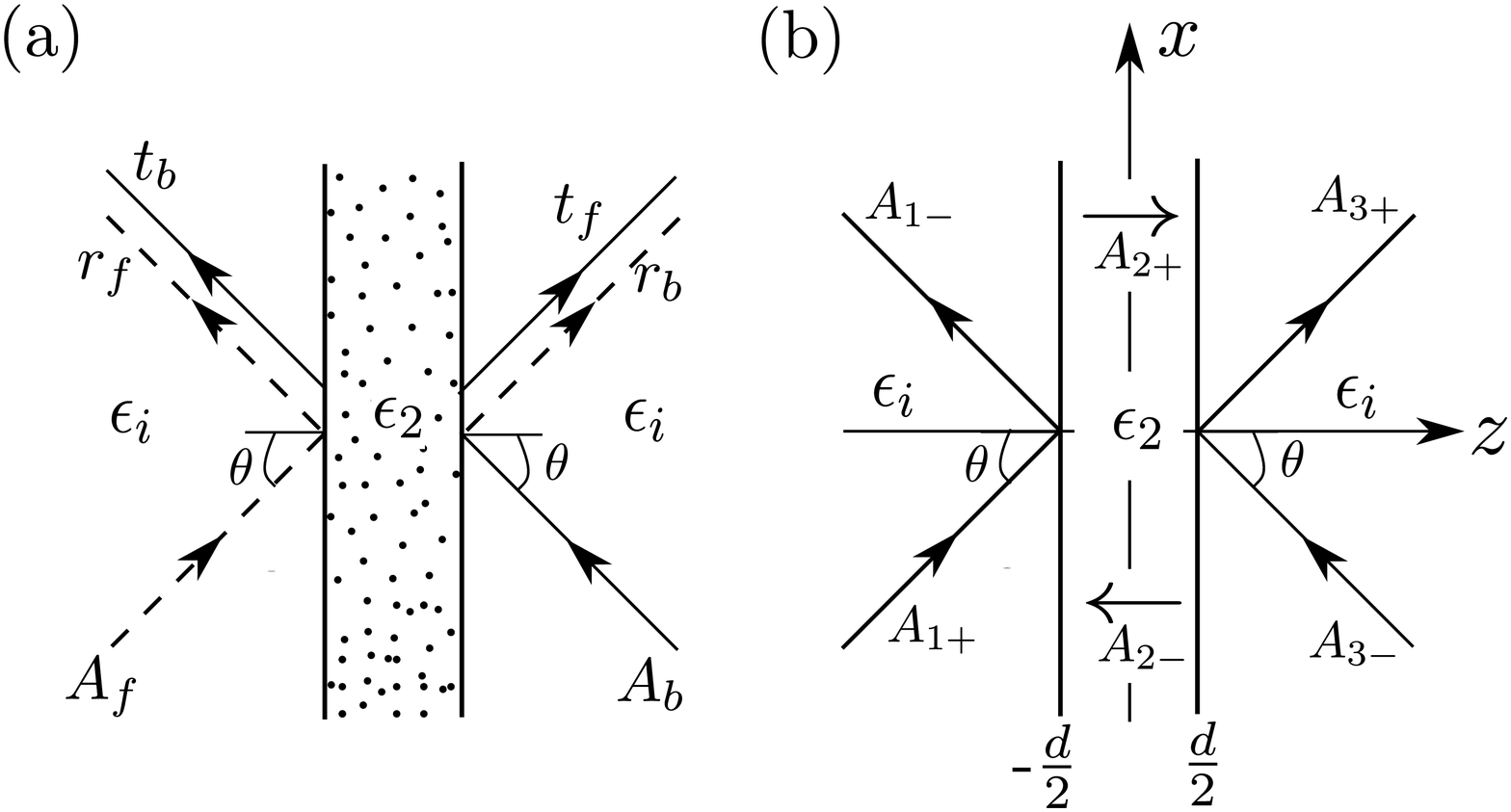}}
\caption{(a) Schematics of the CPA geometry and illumination. (b) Forward and backward propagating waves in different media.} 
\label{fig:2}
\end{figure}
We now briefly explain the mechanism of CPA following Ref.~\cite{opex-12}. Consider the system shown in Fig.~\ref{fig:2}(a), consisting of a slab of width $d$. We assume that the ambient medium are identical in nature to have inversion symmetry. Let the slab be illuminated by coherent monochromatic plane polarized light from opposite ends of the structure symmetrically with unit amplitudes. We denote forward (backward) incident wave amplitude on the left (right) side of the slab and the resulting reflection and transmission by subscript $f$ ($b$). For example, for the forward (backward) incidence the reflected and transmitted amplitudes are given by $r_f$ ($r_b$) and $t_f$ ($t_b$), respectively. By inherent symmetry we have $r_f=r_b$ and $t_f=t_b$ and this implies that the total scattering amplitudes on both the sided are same i.e., $r_f+t_b=r_b+t_f$. The magnitude of intensity scattering depends crucially on the interference effects. For example, the scattered intensity on the left side of the slab is given by $|r_f+t_b|^2$, this can be zero when the two conditions are met i.e., $|r_f|=|t_b|$ and the phase difference between them ($\Delta\varphi=|\varphi_{r_f}-\varphi_{t_b}|$) is an odd multiple of $\pi$. This destructive interference cancels the scattering form the structure on both sides (by the virtue of symmetry) resulting in completing absorption of the incident light known as CPA. Apart from destructive interference one can also have constructive interference when $\Delta\varphi$ is an even multiple of $\pi$ known as super-scattering (SS) \cite{ol-12}. There have been extension of CPA to include the surface modes. Experimental demonstration of CPA mediated by surface plasmons in silver gratings on silicon substrate was given by Yoon \emph{et. al.} \cite{cpa-sp-exp}. There were efforts to make CPA broadband \cite{cpa-brd}. A peculiar property of bending of light on the same side of the normal was shown by exploiting CPA by Shourya \textit{et. al.} \cite{ol-12}. The system consisted of a free standing corrugated metal film, illuminated under CPA geometry, the zeroth order specular reflection was suppressed by CPA leading to effective scattering of light only from the `-1' order, while the `+1' order being in resonance with the surface mode.
\par
In literature one often discusses an interesting parallel of CPA as anti-laser or time reversed laser \cite{trl}. The time reversed lasing action can be understood in the following manner. A laser just above
threshold amplifies the radiation in the active medium and gives out the coherent light in either directions when imperfect mirrors are used. Time reversal in this context amounts to the replacement of gain by loss and reversing the outgoing radiation to incoming ones. Thus, one has complete absorption of the incident radiation by the lossy medium. It
is the destructive interference which nullifies the scattering from the structure if proper phase and amplitude conditions are met. Due to fundamental difference between the optical medium required for lasing (gain) and anti-lasing (loss) actions can not coexist. It was shown by Longhi \cite{longhi-cpa-pt1} that if the medium obeys parity-time ($\mathcal{PT}$) symmetry i.e., $\varepsilon^*(r)=\varepsilon(-r)$ it is possible to have lasing action and CPA in the same medium with appropriate amplitude and phase relations. There have been efforts to extend the ideas from time reversed laser to time reversed surface plasmon amplification by stimulated emission of radiation (SPASER) also \cite{noh,yoon}. From a different angle, there have been investigations to extend the notions of perfect absorption to infrared and terahertz domains in both CC and CPA geometries using graphene \cite{abajo,fan-grap,opex-graph}, as patterned nano-structures of graphene support plasmon resoanances at infrared wavelengths \cite{prb-grap}. 
\par
Apart from the classical notions of perfect absorption, there have been investigations dealing with quantum aspects of radiation \cite{np1,cao-quant,opt-cpa}. It was shown that one can achieve 100\% visibility in Hong-Ou-Mandel dip for two photon quantum interference in a resonant tunneling plasmonic structure \cite{sdg-2p}. This was attributed to perfect destructive interference between the squares of amplitude reflection and transmission coefficients. The idea of CPA was extended to single photons by using path entangled single photons generated by a beam splitter \cite{gsa-cpa}. There have been efforts to understand CPA at a microscopic level also \cite{ieit-gsa}.
\section{Nonlinear critical coupling}\label{S2}
Having discussed the linear properties for CC and CPA, we  review some of the work pertaining to nonlinear systems. The time-reversed lasing action was extended to optical parametric oscillations by Longhi \cite{np-lon-cpa}. The basic idea was to show that the laser gain can be replaced by parametric oscillator gain and the same notions od time reversal still holds and one can realize CPA with colored signal or idler inputs.
CPA in the context of homogeneously broadened two-level medium in an optical cavity was  investigated \cite{np2}. It was shown that because of the dispersive properties of the two-level medium, exact time-reversal symmetry is broken. This results in the difference in the frequencies at which CPA and the lasing mode occur. Furthermore, the time-reversed lasing action has been extended to transient, chaotic, or periodic coherent optical
fields \cite{lon-choas}. CPA devices that perfectly absorb a chaotic laser signal and a frequency-modulated optical wave were also demonstrated.
\par%
Most of the investigations discussed above deal with the extensions of time-reversal aspects of CPA to nonlinear regime. We explore the role of nonlinearity to control CPA/CC resonances. We restrict our attention to Kerr nonlinear stratified medium to probe the possibilities. 
\par%
Our system consists of the layered structure,
 shown in Fig.~\ref{fig:1}. The dielectric function of the absorber layer (a metal-dielectric composite) is modeled using Maxwell-Garnett formula as \cite{Bohren,SDG-CC}
\begin{equation}\label{eq1}
\epsilon_1\left(\omega\right)=\epsilon_h+\frac{fx\left(\epsilon_m-\epsilon_h\right)}{1+f\left(x-1\right)} , \hspace{0.5cm} x=\frac{3\epsilon_h}{\epsilon_m+2\epsilon_h},
\end{equation}
where $f$ is the volume fraction of the inclusion and $\epsilon_m$ ($\epsilon_d$) is the dielectric function of the inclusion (host). The dielectric function of silver is obtained from the proper interpolation of the experimental data of Johnson and Christie \cite{J&C}. It is clear from Eq.~(\ref{eq1}) that for metal [Re$(\epsilon_m)<0$], there can be  resonance when Re$(\epsilon_m)+2\epsilon_d=0$ which is referred to as localized plasmon resonance. Further we assume the spacer layer to be a Kerr nonlinear dielectric given by
\begin{equation}\label{eq2}
\epsilon_2=\tilde{\epsilon}_2+\alpha|E|^2,
\end{equation}
with $E$ as the electric field, $\alpha$ as the nonlinearity constant and $\tilde{\epsilon}_2$ as the linear part of dielectric function. $\alpha>0$ ($\alpha<0$) corresponds to focusing (defocusing) nonlinearity. From now on we consider only the case of focusing nonlinearity. The DBR is made up of  $2N+1$ slabs of alternating `$a$' and `$b$' type layers. The `$a$' (`$b$') type layer is characterized by refractive index $n_a$ ($n_b$) and thickness $d_a$ ($d_b$). Let this structure be illuminated (at normal incidence) by TE polarized monochromatic plane waves of wavelength $\lambda$. The parameters are chosen such that the DBR reflects in the visible range of electromagnetic spectrum and the localized plasmon resonance of the composte medium occurs inside the rejection band of DBR. As discussed earlier, only for certain widths of the spacer layer one satisfies both the conditions for perfect destructive interference at a specific wavelength of light in the stopgap. At this wavelength all the incident light is perfectly absorbed by the structure and this is called critical coupling. The linear CC can found very easily by evaluating the response of the linear system using the standard characteristic matrices \cite{SDG-CC}.%
\par%
We now describe a method to evaluate the nonlinear response of the system. There exists an analytical solution for the nonlinear slab but it is very involved in nature \cite{CM}. A very simple method, namely, nonlinear characteristic matrix theory (NCMT) was developed based on slowly varying envelop approximation (SVEA) \cite{SDG-ncm}. NCMT was used extensively in the past to probe the underlying physics of optical bistability, gap solitons, localization of photons in nonlinear structures, etc \cite{SDG-88,SDG-Imperial,SDG-90,SDG-rev}. We now follow the route of NCMT to evaluate the nonlinear response of the system. The electric field (tangential component) solution to the Maxwell equation inside the nonlinear slab can be written as the superposition of forward and the backward propagating waves with amplitude  dependent phases given by \cite{Felber}
\begin{equation}\label{eq3}
E_{y}=A_+e^{ik_+z}+A_-e^{-ik_-z},
\end{equation}
with $k_{+}$ and $k_{-}$ as the wave vectors corresponding to the forward and backward propagating waves with constant wave amplitudes given by $A_+$ and $A_+$, respectively. $k_{\pm}$ are given by
\begin{equation}\label{eq4}
k_{\pm}=k_{0}\sqrt{\tilde{\epsilon}_2}\left(1+U_{\pm}+2U_{\mp}\right)^{1/2}, 
\end{equation}
with $k_{0}=\omega/c$ and $U_{\pm}=\alpha|A_{\pm}|^2$ as the normalized intensities. Note that $k_{+}$ and $k_{-}$ are different from each other giving rise to nonreciprocity. Starting from the right end (see Fig.~\ref{fig:1}), for a given transmission amplitude $A_t$, the amplitudes in the nonlinear spacer layer are given by \cite{SDG-ncm}
\begin{equation}
\begin{pmatrix} 
  1     & 1\\ 
  \frac{k_+}{k_0} &  -\frac{k_-}{k_0}
\end{pmatrix}\begin{pmatrix} 
  \ A_+    \\ 
  \ A_-  
\end{pmatrix}=
M{_{DBR}}\begin{pmatrix} \ 1 \\ \ \sqrt{\epsilon_f}  \end{pmatrix}A_{t},
\label{eq5}
\end{equation}
where $\sqrt{\epsilon_f}$ is the dielectric constant of the final medium and $M_{DBR}$ is the characteristic matrix of the DBR \cite{bw}. The relations for $U_{\pm}$ can be obtained from Eq.~(\ref{eq5}) as
\begin{eqnarray}
\begin{pmatrix}
U_+ \\ U_-
\end{pmatrix}
= \left|\begin{pmatrix}
1 & 1 \\ \frac{k_+}{k_0} & -\frac{k_-}{k_0}
\end{pmatrix}^{-1}
M_{DBR} \begin{pmatrix}
1 \\ \sqrt{\epsilon_f}
\end{pmatrix} \right|^2 U_t. \label{eq6}
\end{eqnarray} 
In the above equation $U_t=\alpha|A_t|^2$ is the normalized transmitted intensity and $|\cdots|^2$ implies element wise absolute value squared of the vector components. It is evident from Eq.~(\ref{eq4}) that $k_{\pm}$ are functions of $A_{\pm}$ and one doesn't know $A_{\pm}$ in advance to solve Eq.~(\ref{eq6}). But for a given $U_t$,  Eq.~(\ref{eq6}) can be solved by fixed point iteration scheme to yield $k_{\pm}$. Using this information one can now calculate all the elements of nonlinear characteristic matrix \cite{SDG-ncm}. The characteristic matrix for the total structure $M$ can then be written as
\begin{equation} \label{eq7}
M=M_1\times M_2 \times M_{DBR},
\end{equation}
where $M_1$ and $M_2$ denote the characteristic matrices for the absorber and spacer layers, respectively. From the elements of $M$ the intensity reflection ($R$) and transmission ($T$) coefficients can be computed in the usual manner \cite{SDG-rev,bw}. $U_i$ can be calculated as
\begin{equation}
U_i=U_t/T.
\label{eq8}
\end{equation}
The scattering from the structure as function of $U_i$ can then be found by treating $U_t$ as a parameter.
\par%
\begin{figure}[h]
\center{\includegraphics[width=8cm]{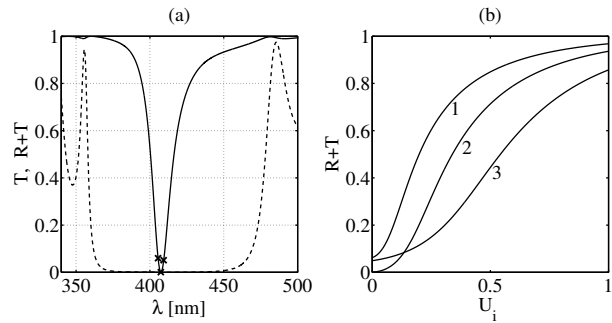}}
\caption{(a) Intensity  transmission coefficient $T$ (dashed curve) and total scattering $R+T$ (solid curve) as functions of the wavelength $\lambda$ for a linear system. (b) $R+T$ as a function of incident intensity $U_{i}$ at three different wavelengths, marked by crosses in Fig.~\ref{fig:3}(a). Corresponding curves are labeled by 1, 2 and 3, respectively, in increasing order of wavelengths. (Plot reproduced from Ref. \cite{knr1}.)}
\label{fig:3}
\end{figure}
We now discuss the numerical results. It is evident form Fig.~\ref{fig:3}(a) that transmission $T$ (dashed curve) is nearly zero due to DBR and dip in the scattering $R+T$ (solid curve) around $\lambda=410~$nm corresponds to CC resonance for the linear structure. We now investigate this at higher power levels to understand the effects of nonlinearity on CC.  Fig.~\ref{fig:3}(b) shows the total scattering $R+T$  as function of input intensity $U_i$ at three different wavelengths (marked by crosses in Fig.~\ref{fig:3}(a)). It is clearly evident from Fig.~\ref{fig:3}(b) that nonlinearity drives the system away from the CC resonance. Note that this is in deep contrast with the Fabry-Perot resonances, which survive even at higher power levels and shift to higher wavelengths by undergoing a nonuniform bending (resulting in multivalued character) \cite{SDG-rev,SDG-ncm}. 
The fundamental difference between optical bistability in Fabry-Perot resonator and the CC resonance can be understood in the following way. Recall that CC demands the perfect balance in amplitude and phase relations (for perfect destructive interference), whereas, in optical bistability nonlinearity amounts to an increase in the effective cavity length and thereby to shifts of the Fabry-Perot resonances.\par%
\par%
\begin{figure}[h]
\center{\includegraphics[width=8cm]{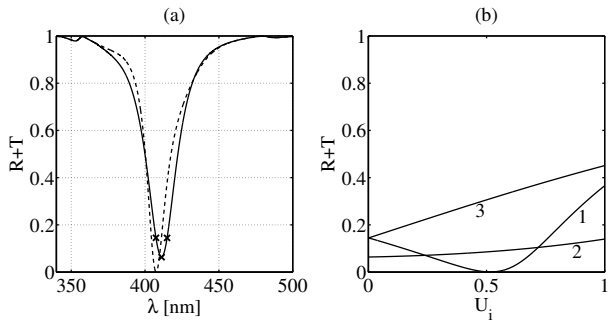}}
\caption{Same in Fig.~\ref{fig:3} but now (a) linear response is shown by solid curve and the dashed curve show the response at higher input intensity $U_i$ (corresponding to the minimum of curve 1 in Fig.~\ref{fig:4}(b)). (Plot reproduced from Ref. \cite{knr1}.)}
\label{fig:4}
\end{figure}
Motivated by the changes induces by nonlinearity, an obvious question can be posed : Can nonlinearity recover CC in the off-resonant system ? We show that CC can be recovered by adjusting the incident power levels. As can be seen from Fig.~\ref{fig:4}(a), a linear system away from the CC resonance (solid curve) can be tuned to CC by increasing the intensity $U_i$ for a specific detuning of the wavelength (see `curve 1' of Fig.~\ref{fig:4}(b)). 
$R+T$ as a function of $\lambda$ at $U_i$ corresponding to the minima of curve 1 [see Fig.~\ref{fig:4}(b)] is shown 
by dashed curve in Fig.~\ref{fig:4}(a). This again confirms the fact that nonlinearity can recover CC in linear off-resonant systems. A system exhibiting dual CC resonances \cite{SDG-CC} is also studied along these lines and we found that one can easily recover CC at one of the wavelengths but it would be difficult to recover CC simultaneously at both the wavelengths.
\section{Nonlinear CPA}\label{S3}
Consider the system shown in Fig.~\ref{fig:2}(b) consisting of a absorbing slab of thickness $d$ illuminated from opposite ends. The dielectric function $\tilde{\epsilon}$ of the gold-silica composite (absorbing slab) is modeled using Bruggeman formulation \cite{opex-12,knr2}. As before, we use the experimental data of Johnson and Christie \cite{J&C} for gold and assume the host (silica) to be dispersionless. We also assume the absorbing slab to be Kerr nonlinear as
\begin{equation}
\epsilon_2=\tilde{\epsilon}+\alpha|E|^2.
\end{equation}
Note that such an inclusion leads to nonlinear absorption, implying complex $\chi^{(3)}$ \cite{tmat,Gehr}. In order to retain the simplicity, we neglect absorption/dispersion in $\chi^{(3)}$ and assume it to be a constant but retain them in $\chi^{(1)}$. Let this slab be illuminated by $s$-polarized monochromatic plane waves. In order to evaluate the nonlinear response of the system, we again follow NCMT \cite{SDG-ncm}. As before, the electric field inside the nonlinear slab can be written as 
\begin{equation}\label{eq10}
E_{y}=\tilde{A}_{2+}e^{ik_{2z+}z}+\tilde{A}_{2-}e^{-ik_{2z-}z},
\end{equation}
with $A_{2+}$ ($A_{2-}$) as the forward (backward) propagating wave amplitude. $k_{2z\pm}$ for oblique incidence are given by
\begin{equation} \label{eq11}
k_{2z\pm}=p_{2z\pm}k_0,~~p_{2z\pm}=\sqrt{\tilde{\epsilon}-p_{x}^2+|A_{2\pm}|^2+2|A_{2\mp}|^2},
\end{equation}
where $p_{2z\pm}$ and $p_{x}=\sqrt{\epsilon_0} \sin \theta$  are normalized propagation constants along $z$ and $x$ directions, respectively, $A_{2\pm}=\alpha|\tilde{A}_{2\pm}|$ are the new dimensionless amplitudes. Evaluating the expression for the tangential component of the magnetic field and rewriting them as elements of column vector and making use of the appropriate boundary conditions at the interfaces at $z=\pm d/2$ gives us
\begin{eqnarray}
\begin{pmatrix}
A_{1+} \\
A_{1-} \end{pmatrix} &=& {\tilde{M}_I}^{-1} \tilde{M}\left(-\bar{d}/2\right)\begin{pmatrix}A_{2+} \\ A_{2-}\end{pmatrix},\label{eq12}\\
\begin{pmatrix}
A_{3+} \\
A_{3-} \end{pmatrix} &=&{\tilde{M}_I}^{-1}\tilde{M}\left(\bar{d}/2\right)\begin{pmatrix}A_{2+} \\ A_{2-}\end{pmatrix}, \label{eq13}
\end{eqnarray}
where $\bar{d}=k_0d$ is the dimensionless width of the slab and  $\tilde{M_I}$ and  $\tilde{M}\left(\bar{d}/2\right)$ are given by 
\begin{widetext}
\begin{equation}
\tilde{M_I}={\begin{pmatrix}
1 & 1 \\
p_{1z} & -p_{1z} \end{pmatrix}},\hspace{0.1cm} \tilde{M}\left(\bar{d}/2\right)=\begin{pmatrix} \exp{(ip_{2z+}\bar{d}/2)} & \exp{(-ip_{2z-}\bar{d}/2)} \\ p_{2z+}\exp{( ip_{2z+}\bar{d}/2)} & -p_{2z-}\exp{(-ip_{2z-}\bar{d}/2)} \end{pmatrix},\label{eq14}  
\end{equation}
\end{widetext}
with $p_{1z}=\sqrt{\epsilon_0-p_x^2}$ as the normalized (to $k_0$) propagating constant in the ambient media. CPA corresponds to $A_{1-}=A_{3+}=0$ i.e, null scattering for finite input. It can be easily shown from Eqs.~(\ref{eq12})-(\ref{eq14}) that, in order to have CPA, the incident intensities must be same ($|A_{1+}|^2=|A_{3-}|^2$). This immediately implies that the field solutions corresponding to CPA in the nonlinear slab can only be either symmetric or antisymmetric in nature. For example, a symmetric (antisymmetric) solution can be realized when $A_{1+}=A_{3-}=A_{in}$ and $A_{2+}=A_{2-}=A_2$ ($A_{1+}=-A_{3-}=-A_{in}$ and $A_{2+}=-A_{2-}=-A_2$). As a consequence, the $z$ component of the normalized propagating constants for forward and backward propagating waves become equal ($p_{2z-}=p_{2z+}=p_{2z}$). Eqs.~(\ref{eq12})-(\ref{eq14}) can then be written as
\begin{eqnarray}\label{eq15}
\begin{pmatrix}
1 & \pm1\\ p_{2z} & \mp p_{2z}
\end{pmatrix} \begin{pmatrix}
A \\ A
\end{pmatrix}&=&M_{d/2}\begin{pmatrix}
1 \\ -p_{1z}
\end{pmatrix}A_{in},
\end{eqnarray} 
where $M_{d/2}$ is the characteristic matrix for the nonlinear half slab. Note that Eq.~(\ref{eq15}) includes the case of linear slab when $\alpha=0$ and $p_{2z}=\sqrt{\tilde{\epsilon}-p_x^2}$. Eq.~(\ref{eq15}) can be further reduced to 
\begin{eqnarray}
D_S=p_{1z}+ip_{2z}\tan{\left(p_{2z}\bar{d}/2\right)}=0,\label{eq16}\\
D_A=p_{1z}-ip_{2z}\cot{\left(p_{2z}\bar{d}/2\right)}=0,\label{eq17}
\end{eqnarray}
for symmetric and antisymmetric solutions, respectively. Solving for $D_S=0~(D_A=0)$ gives us the sufficient conditions for symmetric (antisymmetric) solutions of CPA. The roots of the above equations when solved for complex $\lambda$ yields us the location and the characteristics of CPA dip.\par%
A similar approach can also be employed to find the modes of the Kerr nonlinear waveguide. A waveguide mode corresponds to finite evanescent output for null input ($A_{1+}=0=A_{3-}$). The symmetric (antisymmetric) solutions to waveguiding imply $A_{3+}=A_{1-}=A_{t}$ ($A_{3+}=-A_{1-}=A_{t}$) and Eqs.~(\ref{eq12})-(\ref{eq14}) then become
\begin{eqnarray}\label{eq18}
\begin{pmatrix}
1 & \pm1\\ p_{2z} & \mp p_{2z}
\end{pmatrix} \begin{pmatrix}
A \\ A
\end{pmatrix}&=&M_{d/2}\begin{pmatrix}
1 \\ p_{1z}
\end{pmatrix}A_{t}.
\end{eqnarray} 
Eq.~(\ref{eq18}) can be further reduced to
\begin{eqnarray}
\tilde{p}_{1z}-p_{2z}\tan{\left(p_{2z}\bar{d}/2\right)}=0,\label{eq19}\\
\tilde{p}_{1z}+p_{2z}\cot{\left(p_{2z}\bar{d}/2\right)}=0.\label{eq20}
\end{eqnarray}
for symmetric and antisymmetric modes, respectively, with $\tilde{p}_{1z}=-ip_{1z}$. Eqs.~(\ref{eq19})-(\ref{eq20}) are also referred to as the mode dispersion relations for symmetric waveguide in the literature \cite{burnoski} .\par%
We now draw an interesting parallel to CPA as anti-waveguiding. A comparison between Eqs.~(\ref{eq15}) and (\ref{eq17}) reveals that CPA and waveguiding phenomena form the opposite ends of a scattering events in the following fashion. CPA (waveguiding) corresponds to finite (null) input and null (finite) output. It can be seen from Eqs.~(\ref{eq16})-(\ref{eq17}) that for a linear system, CPA occurs for any input intensities. The nonlinear system behaves completely different as $p_{2z}$ now becomes intensity dependent and CPA can be satisfied only at discrete power levels [due to transcendental character of Eqs.~(\ref{eq16})-(\ref{eq17})].\par%
\begin{figure}[h]
\center{\includegraphics[width=8cm]{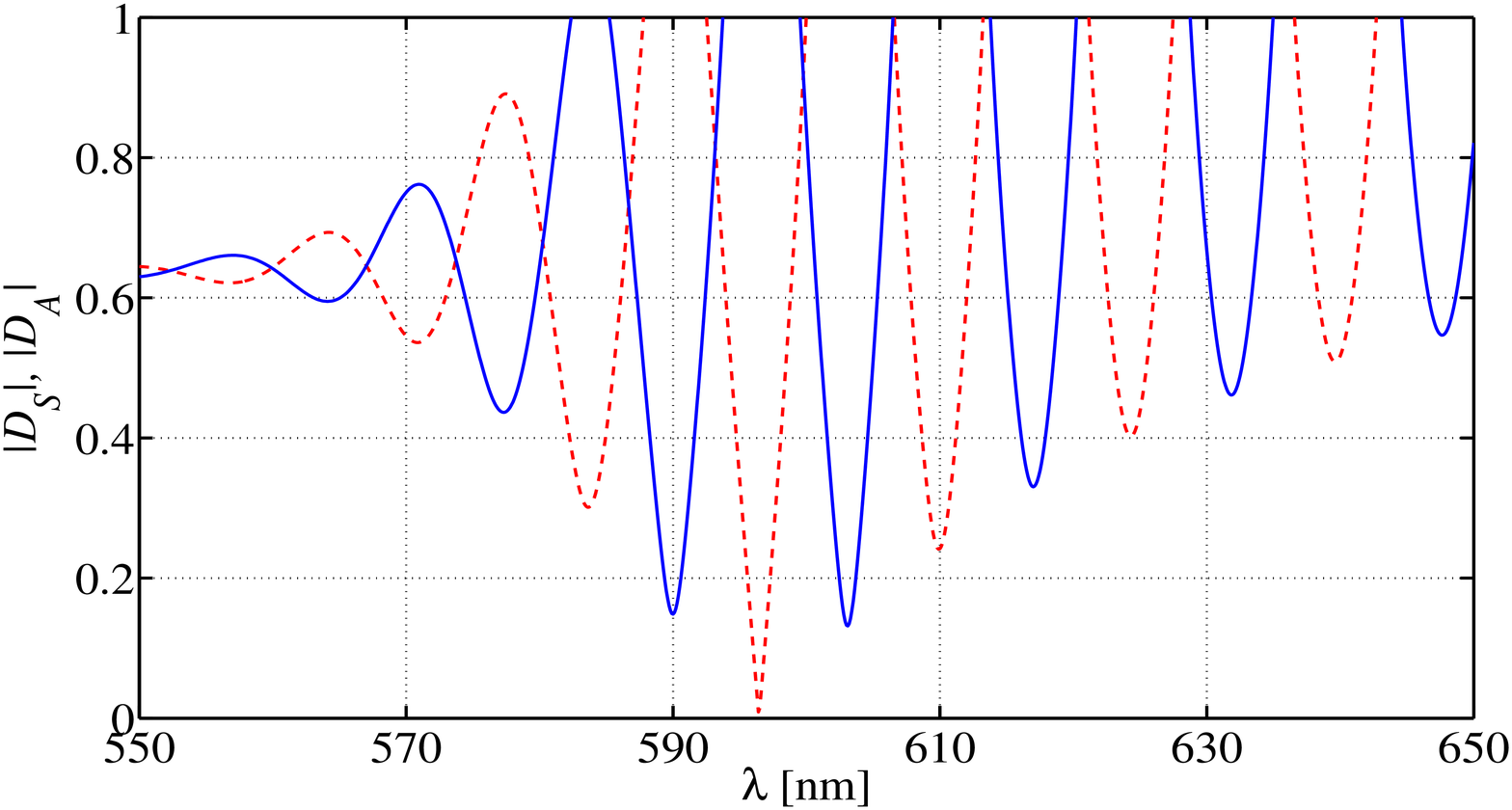}}
\caption{$|D_S|$ (dashed) and $|D_A|$ (solid) as a function of $\lambda$ for linear system. (Plot reproduced from Ref. \cite{knr2}.)}
\label{fig:5}
\end{figure}
\begin{figure}[h]
\center{\includegraphics[width=8cm]{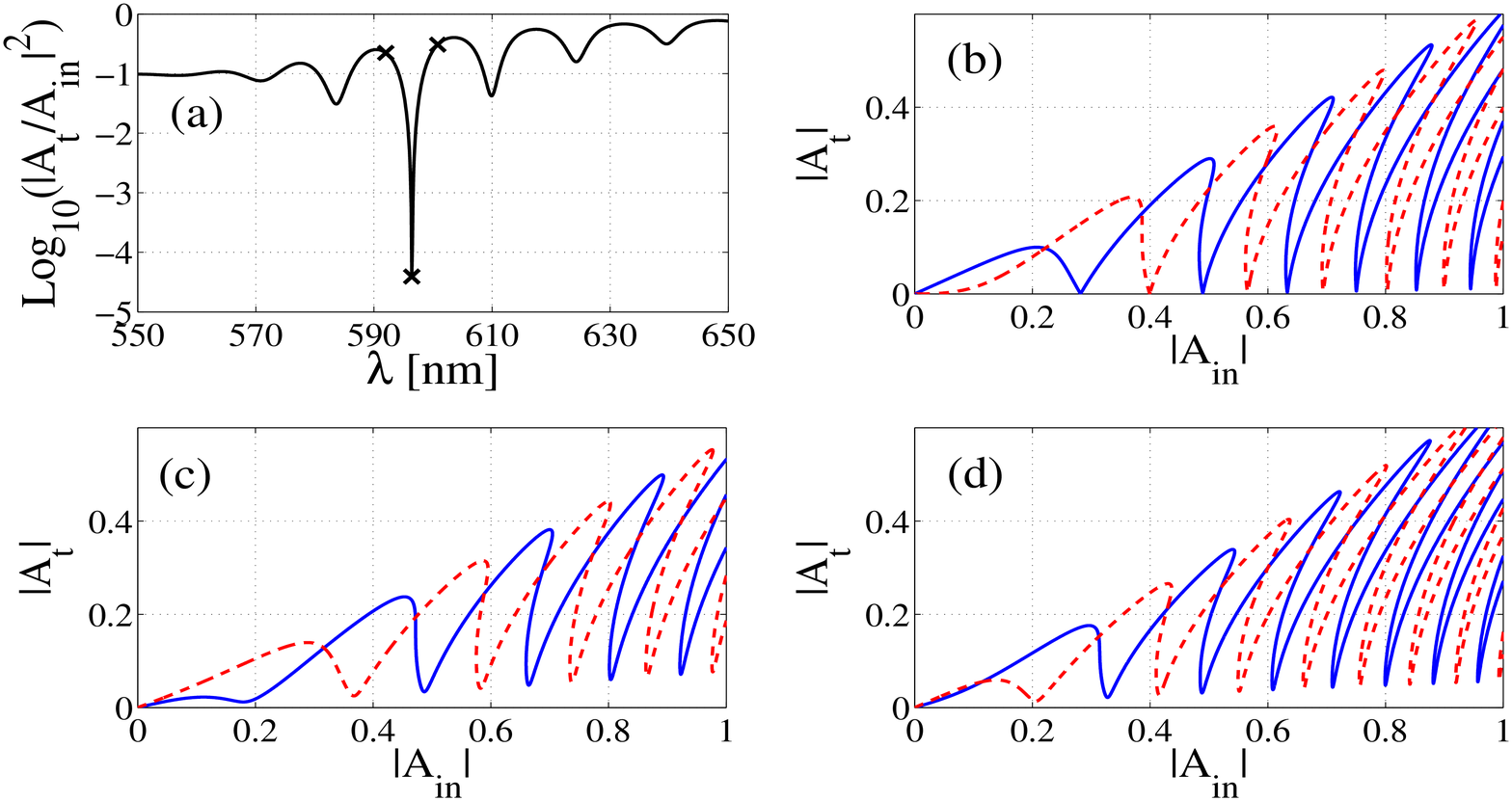}}
\caption{(a) Intensity scattering ($\log_{10}|A_t/A_{in}|^2$) as a function of $\lambda$ for the linear structure. $A_t$ as a function of $A_{in}$ for three different wavelengths marked by crosses, namely, (b) $\lambda=596.4~$nm, (c) $\lambda=592.0~$nm,  (d) $\lambda=600.8$nm for symmetric (dashed) and antisymmetric (solid) states. (Plot reproduced from Ref. \cite{knr2}.)}
\label{fig:6}
\end{figure}

We now present our numerical results and discuss them. We choose to work far away from the plasmon resonance to justify our assumption that $\chi^{(3)}$ is purely a real constant. In order to find the location of CPA in linear system we plot $|D_S|$ (dashed curve) $|D_A|$ (solid curve) as a function of $\lambda$. It is evident from Fig.~\ref{fig:5} that only $D_S=0$ has a root at $\lambda=596.4~$nm and $D_A=0$ doesn't have any root in this range. Thus, CPA in linear system occurs at $\lambda=596.4~$nm and it is symmetric in nature. To confirm this we plot $|A_{t}/A_{in}|^2$ (scattering) as a function of $\lambda$ in Fig.~\ref{fig:6}(a) for the same system. The near-null dip in the scattering occurs exactly at the same $\lambda$.\par
\begin{figure}[h]
\center{\includegraphics[width=8cm]{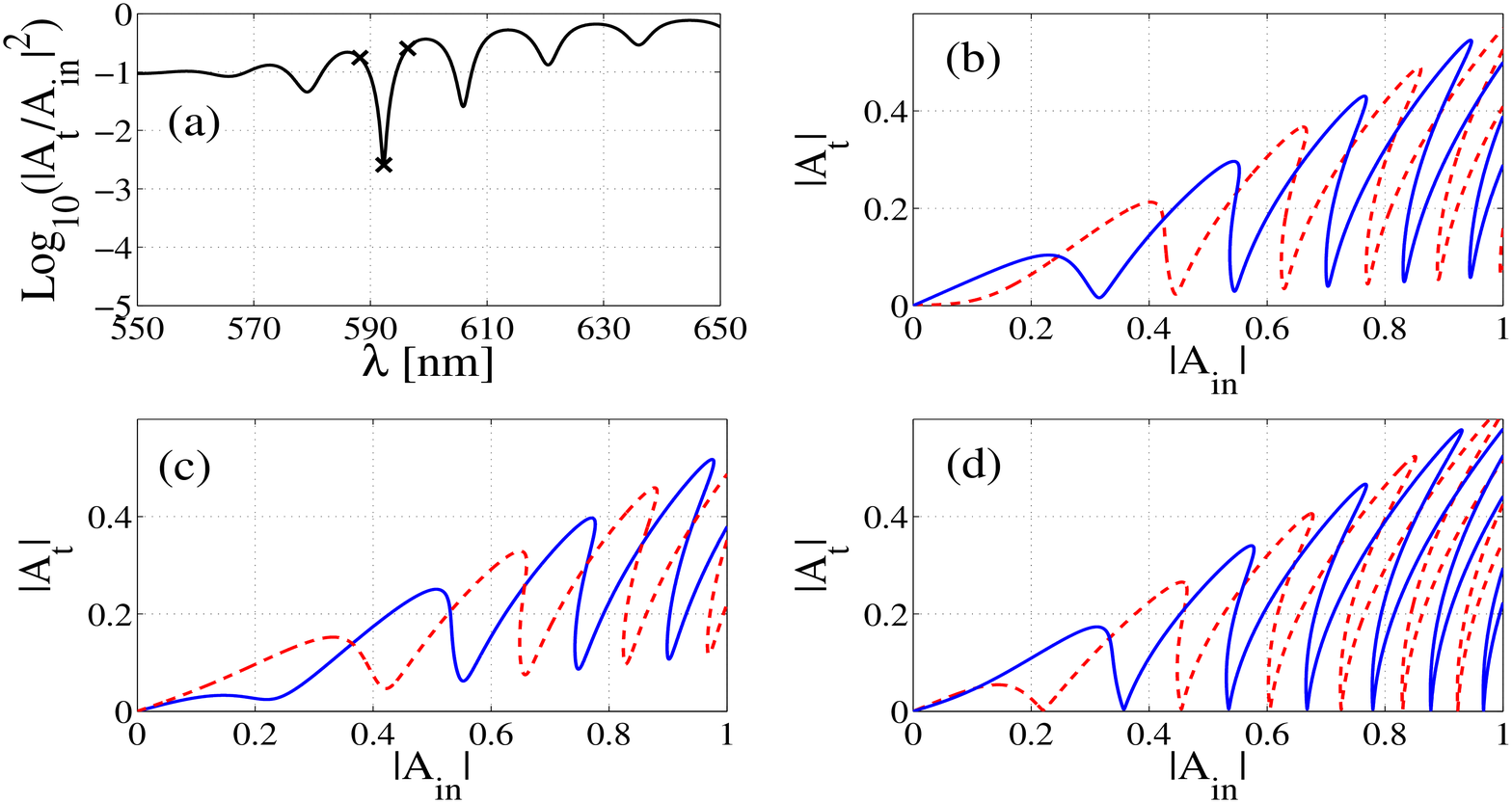}}
\caption{Same as in Fig.~\ref{fig:6}, but now for the linear off-resonant system with (a) $d=18.50~\mu$m (b) $\lambda=592.3$, (c) $\lambda=588.2~\mu$m, (d) $\lambda=596.4~\mu$m. (Plot reproduced from Ref. \cite{knr2}.) }
\label{fig:7}
\end{figure}

%
The nonlinear response of the system can be found by treating $A$ as a parameter, and relating $A_{t}$ with $A_{in}$. Thus, CPA would mean $A_{t}$ going to zero. As before, we study the effects of nonlinearity on a linear on-resonant system. Figs.~\ref{fig:6}(b)-(d) show the nonlinear response of the system at three different wavelengths marked by crosses in Fig.~\ref{fig:6}(a). We find that only for the wavelength tuned at the resonance (see Fig.~\ref{fig:6}(b)) one can achieve CPA even at discrete power levels, whereas, any other wavelength detuning fails to do so  (see Figs.~\ref{fig:6}(c)-(d)). It is again evident from Fig.~\ref{fig:6}(b) that CPA at very low power levels (linear regime) is again symmetric in nature which is consistent with Fig.~\ref{fig:5}. We also demonstrate that one can recover CPA in the linear off-resonant system by tuning the power incident power levels. Fig.~\ref{fig:7}(a) corresponds to the response of the linear off-resonant system. Fig.~\ref{fig:7}(b)-(d) show the nonlinear response of at three different wavelengths marked by crosses in Fig.~\ref{fig:7}(a). It can be seen from Fig.~\ref{fig:7}(b) only for the detuning on the right of the dip of Fig.~\ref{fig:7}(a) one can recover CPA at higher power levels. Note the multivalued character of $A_t$ for a given $A_{in}$ in the nonlinear response of Figs.~\ref{fig:6} and \ref{fig:7} which can be very useful for switching devices. %
As mentioned earlier, CPA (SS) corresponds to constructive (destructive) interference. One can switch from CPA to super-scattering by introducing an additional phase difference of $\pi$ between the incident beams. The symmetric ($A_{1+}=A_{3-}$) and antisymmetric ($A_{1+}=-A_{3-}$) excitation scheme differ only by a phase difference of $\pi$. From this one finds that the CPA of symmetric solution corresponds the SS of antisymmetric solution and vise versa. It can be seen from the nonlinear response of Figs.~\ref{fig:6} and \ref{fig:7} that the dip of symmetric solution doesn't correspond to the peak of the antisymmetric solution. This difference can be attributed to the changes introduced by nonlinearity.
\section{CPA with gap solitons}\label{S4}
In this section, we further extend the investigations from a single slab to a periodic layered medium. We consider a symmetric periodic layered medium with alternating Kerr nonlinear and linear dielectrics illuminated in CPA-like geometry. In the context of one sided illumination It was shown that the nonlinearity induced total transmission states correspond to soliton-like spatial profiles known as gap solitons \cite{CM-gap,Sipe}. In addition to the case of normal incidence, gap solitons can be excited at oblique incidence \cite{SDG-Imperial}.
 \par%
In the present context, we ask the question: Is it possible to channel all the incident light energy into these gap soliton modes ? We show that for a symmetric periodic structure illuminated from opposite ends one can excite gap solitons with null scattering from the structure, resulting in complete transfer of energy into the gap soliton modes.
\begin{figure}[h]
\center{\includegraphics[width=8cm]{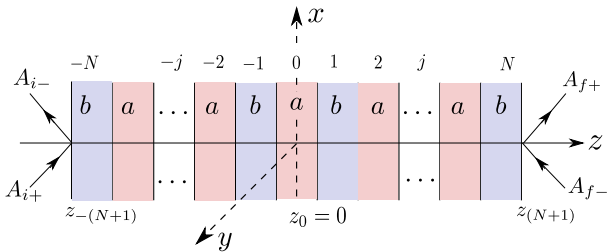}}
\caption{Schematics of periodic layered media and illumination. The `$a$' (`$b$') type layers are nonlinear (linear) with dielectric function $\epsilon_a$ ($\epsilon_b$) and width $d_a$ ($d_b$). The layers are labeled from the center with even (odd) integer corresponding to the nonlinear (linear) layers.}
\label{fig:8}
\end{figure}
We consider the system shown in Fig.~\ref{fig:8}, with $a$ ($b$) type layer as Kerr nonlinear (linear) dielectric layer. The total number of periods (with each period composed of linear and a nonlinear layer) are assumed to be $N$ resulting in $2N+1$ layers. We further assume that the central layer is nonlinear and the extreme layers are linear. Let this structure be illuminated by TE polarized monochromatic plane waves of wavelength of $\lambda$, at an angle $\theta$. Maxwell's equations for the tangential components of the electric and magnetic fields in any $j^{th}$ layer are given by
\begin{eqnarray} \label{eq21}
\frac{dE_{jy}}{dz}&=&-iH_{jx},\\ 
\frac{dH_{jx}}{dz} &=&\begin{cases} \label{eq22}
       -i[(\epsilon_j-p_{x}^2)+|E_{jy}|^2]E_{jy} \hspace{0.15cm} \text{for even $j$,}\\ 
       -i[(\epsilon_j-p_{x}^2)]E_{jy} \hspace{0.15cm} \text{for odd $j$}, 
     \end{cases} 
\end{eqnarray}
with $p_x=\sqrt{\epsilon_i}\sin \theta$. Eqs.~(\ref{eq21})-(\ref{eq22}) are written in terms of dimensionless variables given by $z\rightarrow k_0z,~E\rightarrow\sqrt{\alpha}E$. Note that $E_{jy}$ and $H_{jx}$ are in general, complex resulting in four coupled differential equations for real and imaginary parts. This set of coupled differential equations can be solved numerically exactly (see below) and also approximately. \par%
We now outline a method to solve Eqs.~(\ref{eq21})-(\ref{eq22}) in a approximate method using NCMT \cite{SDG-Imperial,SDG-ncm}. As before, the electric field in any $j^{th}$ layer can be expressed as
\begin{equation}
E_{jy}=A_{j+}e^{ip_{jz+}(z-z_j)}+A_{j-}e^{-ip_{jz-}(z-z_j)}, \label{eq23}
\end{equation} 
with $z_j \leq z \leq z_{j+1},~z_0=0$ and $p_{jz\pm}$ are given by
\begin{equation}
p_{jz\pm}=\begin{cases}
\sqrt{\epsilon_j-p_x^2+(|A_{j\pm}|+2|A_{j\mp}|)} \hspace{0.2cm} \text{for even $j$}, \\ 
\sqrt{\epsilon_j-p_x^2}\hspace{0.2cm} \text{for odd $j$}. 
\end{cases}
\label{eq24}
\end{equation} 
In order to ensure that waves are propagating (evanescent) in nonlinear (linear) layers, we restrict $\theta$ in the range
\begin{equation}
\epsilon_a-p_x^2>0, \hspace{0.5cm} \epsilon_b-p_x^2<0,\label{eq25}
\end{equation}
so that the structure shown in Fig.~\ref{fig:8} represents a system of $N$ coupled nonlinear waveguides, coupled evanescently. Thus, the structure sandwiched between two high index prisms represents a resonant tunneling geometry \cite{RT1,RT2}. When the resonant modes (also called supermodes) are supported by the structure, one has complete transmission from the structure \cite{yeh,SDG-Imperial}.\par%
As before, exploiting the symmetry in the central slab it can be shown that solutions to CPA in the central layer can only be either symmetric ($A_{0+}=A_{0-}=A_0$) or antisymmetric ($A_{0+}=-A_{0-}=-A_0$) in nature (see Sec.~\ref{S2}).  Starting from the central layer with $A_0$ as a parameter, one can find the  incident amplitude $A_{i+}$ ($A_{f-}$) and the scattered amplitude $A_{i-}$ ($A_{f-}$) on the left (right) side outside the layered medium. For example, starting from $z_0=0$ in the central layer, the amplitudes ($A_{-2\pm}$) in the nonlinear layer labeled $j=-2$ are given by
\begin{eqnarray}
\begin{pmatrix}
A_{-2+} \\ A_{-2-} \end{pmatrix} = \begin{pmatrix}  1 & 1 \\ p_{-2z+} & -p_{-2z-} \end{pmatrix}^{-1}M_1(d_b)\times \nonumber \\ \times M_0(d_a/2)\begin{pmatrix}  1 & 1 \\ p_{0z} & -p_{0z} \end{pmatrix}\begin{pmatrix}A_0 \\ \pm A_0 \end{pmatrix}, \label{eq26}
\end{eqnarray}
with $M_1(d_a/2)$ and $M_2(d_b)$ as the characteristic matrices for nonlinear and linear layers of widths $d_a/2$, $d_b$, respectively. Since Eq.~(\ref{eq26}) is a nonlinear equation as Eq.~(\ref{eq5}), we solve them as outlined in Sec.~\ref{S2}. Repeating the same procedure for all the layers till end, for example, on the left extreme we have
\begin{eqnarray}
\begin{pmatrix}
A_{i+} \\ A_{i-} \end{pmatrix} = \begin{pmatrix}  1 & 1 \\ p_{iz} & -p_{iz} \end{pmatrix}^{-1} M_N(d_b) M_{N-1}(d_a) \cdots \nonumber \\ \cdots M_1(d_b)M_0(d_a/2)\begin{pmatrix}  1 & 1 \\ p_{0z} & -p_{0z} \end{pmatrix}\begin{pmatrix}A_0 \\ \pm A_0 \end{pmatrix}, \label{eq27}
\end{eqnarray}  
with $p_{iz}$ as the normalized propagation constant of the incident ($A_{i+}$) and scattered ($A_{i-}$) amplitudes on the left side of the structure. The scattered (incident) amplitudes $A_{f+}$ ($A_{f-}$) on the right side of the structure can also found by propagating towards right from the center. As a consequence of inherent symmetry (about $z_0=0$ ), the symmetric (antisymmetric) solutions correspond to $A_{i+}=A_{f-}=A_{in}$ ($A_{i+}=-A_{f-}=-A_{in}$), and $A_{i-}=A_{f+}=A_{t}$ ($A_{i-}=-A_{f+}=-A_{t}$). We then define normalized scattering intensity $S=|A_{i-}/A_{i+}|^2=|A_{f+}/A_{f-}|^2=|A_{t}/A_{in}|^2$. For nonlinear response, we plot $S$ as a function of $U_{in}=|A_{in}|^2$, $S\rightarrow0$ would correspond to CPA.  \par%
We now exploit symmetry to solve Eqs.~(\ref{eq21})-(\ref{eq22}) exactly numerically. The solution to electric and magnetic fields in the central layer under CPA-like illumination scheme are either symmetric or antisymmetric with
\begin{eqnarray}
E_y(-z)&=E_y(z),\hspace{0.5cm}H_x(-z)=-H_x(z),&\\ \label{eq28}
E_y(-z)&=-E_y(z),\hspace{0.5cm}H_x(-z)=H_x(z),&  \label{eq29}
\end{eqnarray}
respectively. Thus at $z_0=0$ in the central layer we have $H_x=0$ ($E_y=0$) for the symmetric (antisymmetric) solutions. The nonvanishing components $E_y$ ($H_x$) at $z_0=0$ for symmetric (antisymmetric) solution is taken as a parameter to find to the normalized scattering intensity $S$ outside the structure in the following way. Eqs.~(\ref{eq21})-(\ref{eq22}) can be integrated numerically (with two initial conditions given by $E_y$ and $H_x$ at $z_0=0$) until the left/right end ensuring the continuity of the fields at every interface. As the ambient media are assumed to be linear, the incident and the scattered amplitudes can be computed in the usual way.\par%
\begin{figure}[h]
\center{\includegraphics[width=8cm]{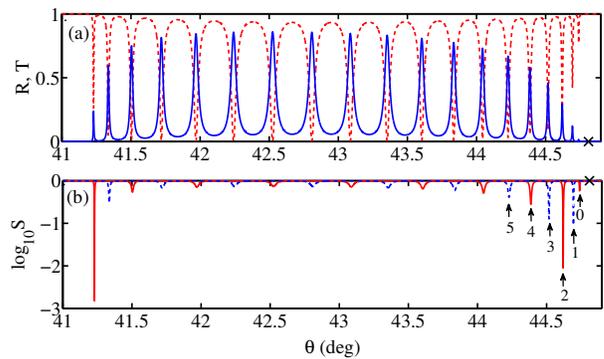}}
\caption{Linear response (a) $R$ (dashed) and $T$ (solid) as a function of $\theta$ for unidirectional illumination; (b) normalized intensity scattering $S$ in $\log_{10}$ scale as a function of $\theta$ for bidirectional illumination with solid and dashed curves corresponding to symmetric and antisymmetric solutions, respectively. The CPA-like dips are labeled from the right edge with even (odd) integers for symmetric (antisymmetric) cases. (Plot reproduced from Ref. \cite{knr3}.)}
\label{fig:9}
\end{figure}
We now present the numerical results. We choose $\theta$ in the range $41^{\circ}$ to $45^{\circ}$ to satisfy the condition in Eq.~(\ref{eq25}). Most of the results presented are exact, obtained by numerically integrated Eqs.~(\ref{eq21})-(\ref{eq22}). We also present some of the results obtained by approximate method in order to compare them with the exact ones.\par%
We first present the linear results and discuss them. $R$ and $T$ as a function of $\theta$ for unidirectional illumination are shown in Fig.~\ref{fig:9}(a). The sharp resonances in the small range of $\theta$ are due resonant tunneling of electromagnetic radiation. The number of such resonances exactly coincides with the number of waveguides in the structure. At every resonance the phase difference between ($\Delta\varphi=|\varphi_r-\varphi_t|$) the reflected and the transmitted light undergoes a characteristic phase jump of $\pi$. If now one satisfies the equality $|r|=|t|$  near any of the resonances, for a bidirectional illumination this would mean CPA.  We have plotted the normalized intensity scattering $S$ from the structure for both symmetric and antisymmetric solutions for bidirectional illumination in Fig.~\ref{fig:9}(b) to investigate linear CPA.
\begin{figure}[h]
\center{\includegraphics[width=8cm]{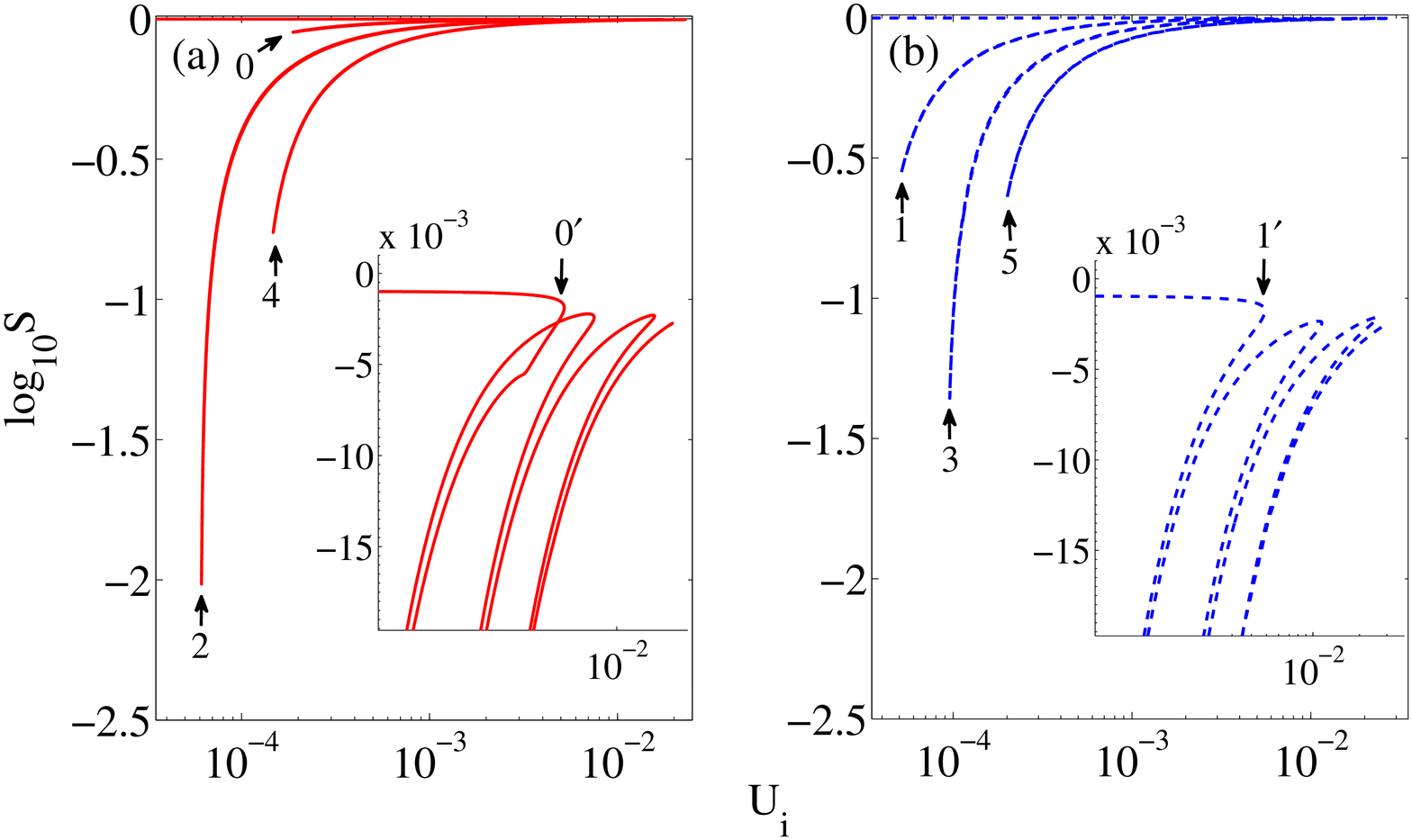}}
\caption{Nonlinear response for bidirectional illumination. Normalized intensity scattering $\log_{10}S$ as a function of incident intensity $U_i$ for (a) symmetric and (b) antisymmetric solutions with insets showing the corresponding nonlinear SS states. The dips are labeled following Fig.~\ref{fig:9}. (Plot reproduced from Ref. \cite{knr3}.)}
\label{fig:10}
\end{figure}
The CPA-like dips can be seen from Fig.~\ref{fig:9}(b), these are labeled from the extreme right end by the integers 0, 2, 4 (1, 3, 5) for symmetric (antisymmetric) states. A prominent CPA-like dip can be seen for the curve labeled 2 [see Fig.~\ref{fig:9}(b)] as the both conditions for perfect destructive interference are met i.e., $|r|\approx|t|$ and phase difference of $\pi$ due the characteristic phase jump at the resonance.\par
We now investigate the nonlinearity induced changes in the stopgap of the linear structure by increasing the incident power levels. It is evident from Fig.~\ref{fig:9}(a) that the stopgap of the linear structure begins at $\theta=44.76^{\circ}$ on the right end. We chose $\theta=44.81^{\circ}$ [a point marked by cross in Fig.~\ref{fig:9}] inside the stopgap as our operating point. For nonlinear response we plot $S$ as a function of $U_i$ for both symmetric and antisymmetric solution in Fig.~\ref{fig:10}(a) and \ref{fig:10}(b), respectively. Fig.~\ref{fig:10} clearly demonstrates that CPA can be realized even in the stop at higher incident power levels. The CPA-like dips are labeled following their linear counterparts in Fig.~\ref{fig:9}(b). It may be noted that, since nonlinearity drives the on-resonant CPA system away from CPA, it would be difficult to have prominent CPA dips both in the linear as well as nonlinear regimes. The insets of Figs.~\ref{fig:10}(a) and \ref{fig:10}(b) clearly demonstrate the nonlinear SS states for symmetric and antisymmetric states, respectively. The increase in incident intensity would correspond to extra optical path (for focusing nonlinearity) leading to a shift of CPA-like dips towards the right edge (see Fig.~\ref{fig:9}(b)). These dips undergo a nonuniform bending resulting in multivalued character as shown in Fig.~\ref{fig:10}.\par%
We also computed the electric field intensity $|E|^2$ in the total structure corresponding to the CPA minima and SS maxima using both exact and the approximate methods. 
\begin{figure}[h]
\center{\includegraphics[width=7cm]{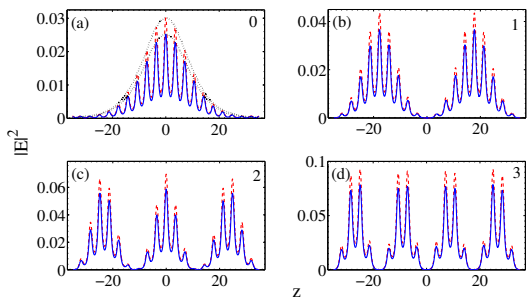}}
\caption{$|E|^2$ inside the total structure as a function of $z$ corresponding the nonlinear CPA-like dips shown in Fig.~\ref{fig:10}. Solid (dashed) curves depict the exact (approximate) solutions. Panels of Fig.~\ref{fig:11}(a)-(d) correspond to CPA-like dips labeled 0-3, respectively. The dotted line in Fig.~\ref{fig:11}(a) shows the envelope fitted with $A/\cosh^2 (\beta z)$. (Plot reproduced from Ref. \cite{knr3}.)}
\label{fig:11}
\end{figure}%

\begin{figure}[h]
\center{\includegraphics[width=7cm]{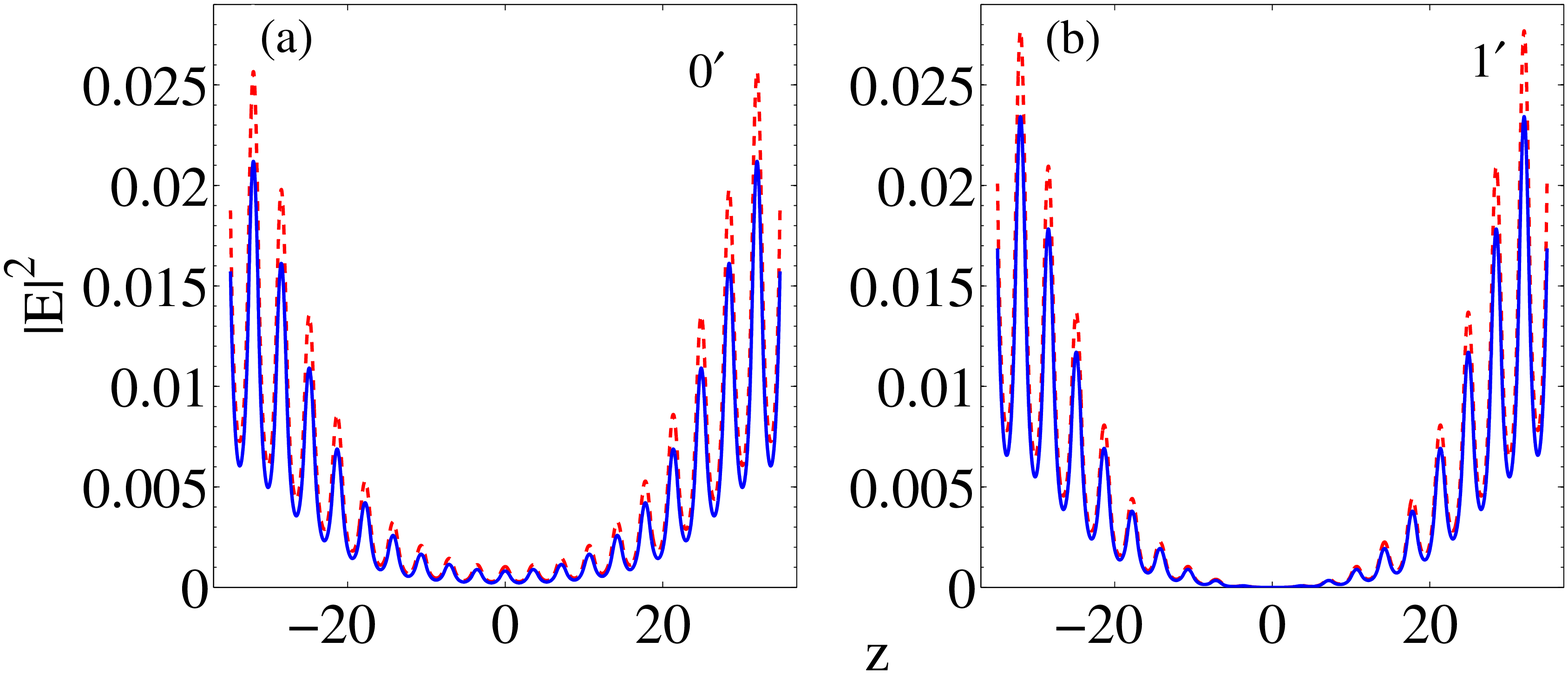}}
\caption{Same as in Fig.~\ref{fig:11} but now for nonlinear SS states (see insets of Fig.~\ref{fig:10}) corresponding to (a) symmetric ($0^{\prime}$) and (b) antisymmetric ($1^{\prime}$) solutions. (Plot reproduced from Ref. \cite{knr3}.)}
\label{fig:12}
\end{figure}

 The typical field profiles corresponding to the symmetric and antisymmetric CPA-like dips are shown in fig.~\ref{fig:11}. One can see one-, two-, many-, soliton-like intensity profiles corresponding to CPA-like dips. For example, the field profile corresponding to the CPA-like dip labeled by `0' shown in Fig.~\ref{fig:11}(a) can be fitted with $A/\cosh^2 (\beta z)$ for both exact and approximate methods. Unlike the CPA case, the lower order SS intensity profiles are localized away from the center of the structure (see Fig.~\ref{fig:12}).\par%
It can can seen from Figs.~\ref{fig:11} and \ref{fig:12} that the approximate and the exact methods differ very little with approximate method estimating the intensities slightly higher than that of exact one. This good agreement validates the SVEA in the present context. 
\section{Conclusions}\label{S5}
In conclusion, we have presented a brief review of linear and nonlinear effects in systems supporting CC and CPA. We have shown how laser power plays an important role in recovering the CC and CPA dips in detuned systems. Further, we demonstrate recovery of CPA at discrete values of incident powers. Applied to a nonlinear periodic structure, gap solitons are shown to exists which do not scatter any light. Results reported here can find many applications in linear and nonlinear optical devices.\par%
\vskip2ex 
One of the authors (SDG) gratefully acknowledges the invitation for writing this review.
\bibliographystyle{ieeetr}
\bibliography{nireek}

\end{document}